# Magnetic droplet nucleation with homochiral Néel domain wall

Sanghoon Kim[1†], Peong-Hwa Jang[2], Duck-Ho Kim[1], Takuya Taniguchi[1], Takahiro Moriyama[1], Kab-Jin Kim[1,4], Kyung-Jin Lee[2,3], and Teruo Ono[1†]

[1]*Institute for Chemical Research, Kyoto University, Gokasho, Uji, Kyoto 611-0011, Japan*

[2]*Department of Materials Science & Engineering, Korea University, Seoul 02841, Korea*

[3]*KU-KIST Graduate School of Converging Science and Technology, Korea University, Seoul 02841, Korea*

[4]*Department of Physics, Korea Advanced Institute of Science and Technology, Daejeon 34141, Korea*

† Correspondence to : makuny80@gmail.com, ono@scl.kyoto-u.ac.jp

**Abstract**

We investigate the effect of the Dzyaloshinskii–Moriya interaction (DMI) on magnetic domain nucleation in a ferromagnetic thin film with perpendicular magnetic anisotropy. We propose an extended droplet model to determine the nucleation field as a function of the in-plane field. The model can explain the experimentally observed nucleation in a Co/Ni microstrip with the interfacial DMI. The results are also reproduced by micromagnetic simulation based on the string model. The electrical measurement method proposed in this study can be widely used to quantitatively determine the DMI energy density.

Chiral spin textures, such as a skyrmion and a Néel type magnetic domain wall (DW), which are formed in perpendicularly magnetized ferromagnet (FM)/non-magnet (NM) structures, have attracted considerable attention for future spintronic devices that can perform massive and high-speed data processing with low power consumption [1-7]. Lack of structural inversion symmetry with strong spin-orbit coupling at the FM/NM interface results in the Dzyaloshinskii–Moriya interaction (DMI) which results in chirality in spin textures [8-10]. In recent years, it has been experimentally demonstrated that skyrmion-like bubble domains can be formed stably at room temperature [11-12] and driven along a magnetic nanowire at a velocity of ~100 m/s [13,14]. In addition, a homochiral Nèel type DW is stabilized by the DMI and is driven by current [15-19] at a velocity of as high as 750 m/s. The DMI can influence the dynamical properties of topological spin textures [20], as proven in dynamic DW measurements where the DMI suppresses DW velocity breakdown [20]. Moreover, the DMI can affect current-driven switching and the diode effect of a magnetic tunnel junction [21-24]. Therefore, the mechanism of formation, reversal, and modulation of chiral magnetic objects should be investigated to understand the characteristics of spin-based devices with DMI.

In this study, we examine the effect of the DMI on magnetic domain nucleation, which is the starting point for magnetic object formation or reversal in an FM/NM bilayer. First, we extended the magnetic droplet model by including the DMI effect. Considering a one-dimensional magnetization configuration, the extended model predicts that the nucleation field ($H_n$, parallel to out-of-plane direction) of a droplet in an FM/NM bilayer is affected by the application of an in-plane bias field ($H_{in}$). For small $H_{in}$, $H_n$ is almost constant; however, for large $H_{in}$, $H_n$ decreases with increasing $H_{in}$. The threshold value of $H_{in}$ originates from the DW energy of the magnetic droplet and is strongly dependent on the DMI. Therefore, the measurement of the threshold behavior of $H_{in}$ enables us to quantify the DMI. We validate the model through angle-dependent nucleation field measurement and micromagnetic simulation using Co/Ni multilayer systems. The results of the experiment and simulation are in good agreement with calculation results using the extended droplet model, suggesting that the model provides a method to quantify $D$ using general anomalous Hall effect (AHE) measurement. Even though chirality-induced magnetic nucleation has been reported recently [25], the proposed analytical approach explains the characteristics of a full-circle droplet, such as threshold behavior under $H_{in}$, which has been an unresolved question.

For a magnetic droplet in a perpendicularly magnetized medium, $H_n$ is determined by the Zeeman energy and total DW energy ($\sigma_{DW,total}$) of the magnetic droplet as

$$H_n = \frac{\pi \sigma_{DW,total}^2 t}{2\mu_0 M_S p k_B T}, \qquad (1)$$

where $t$ is the film thickness, $\mu_0 M_S$ is the saturation magnetization, $k_B$ is the Boltzmann constant, $T$ is the temperature, $p$ is the thermal stability factor related to the activation time, $\tau$ ($=\tau_0 e^p$), for the nucleation of a reversed domain, and $\tau_0$ is the attempt frequency [26]. We note that $H_n$ is a quadratic function of $\sigma_{DW}$. As $\sigma_{DW}$ is modulated by the DMI and $H_{in}$ [25], the measurement of $H_n$ as a function of $H_{in}$ allows for estimation of the DMI.

We extend the droplet model to include the effect of $H_{in}$ and the DMI on $\sigma_{DW}$. When the DMI is stronger than DW anisotropy energy ($K_D$), homochiral Nèel type DWs are formed. Here, we assume that two DW magnetizations, parallel and antiparallel to $H_{in}$, dominate the $\sigma_{DW}$ of the droplet [see white arrows in Fig. 1(a)]. When $H_{in}$ (=$H_x$) is applied in the $x$ direction, the total $\sigma_{DW}$ is given by $\sigma_{DW,total}(|H_x|) = \sigma_{DW1}(+H_x) + \sigma_{DW2}(-H_x)$. The positive and negative signs of $H_x$ are required for $\sigma_{DW1}$ and $\sigma_{DW2}$, respectively, because the magnetizations of two small DW sections ($M_{DW1}$ and $M_{DW2}$) are approximately anti-parallel [see white arrows in Fig. 1(a)], i.e., DW1 and DW2 are parallel and antiparallel to the direction of $H_x$, respectively. Owing to the chirality resulting from the DMI, the $\sigma_{DW}$s of both DW magnetizations should be different under $H_x$. Here, the $\sigma_{DW}$ with a single magnetization direction is given by [27]

$$\sigma_{DW}(H_x) = \begin{cases} \sigma_0 - \dfrac{\pi^2 \Delta M_S^2 \mu_0^2}{8K_D}(H_x + H_{DMI})^2 & , \text{ for } \mu_0|H_x + H_{DMI}| < \dfrac{4K_D}{\pi M_S} \\ \sigma_0 + 2K_D\Delta - \pi\Delta M_S \mu_0 |H_x + H_{DMI}| & \text{otherwise,} \end{cases}$$

(2)

where $\sigma_0$ (=$4\sqrt{AK_{eff}}$) is the energy of a Bloch-type DW, $A$ is the exchange stiffness constant, $K_{eff}$ is the effective perpendicular anisotropy energy, and $\Delta$ (=$\sqrt{A/K_{eff}}$) is the DW width. Here, $4K_D/\pi M_S$ is the magnetic field strength required to saturate the magnetization in the DW of the droplet with $H_x$. For $|H_x| = 0$, $\sigma_{DW,total}(0) = 2\sigma_{DW}(0)$. When $|H_x| > |H_{DMI}|$, $\sigma_{DW,total}$ is given by

$$\sigma_{DW,total}(|H_x|) = \sigma_{DW1}(+H_x) + \sigma_{DW2}(-H_x)$$

$$= 2\sigma_0 + 4K_D\Delta - 2\pi\Delta M_S\mu_0 H_x. \quad (3)$$

Figure 1(b) shows the $\sigma_{DW,total}^2$ vs. $H_x$ curves in terms of $H_{DMI}$, calculated based on the $\sigma_{\text{DW,total}}$ discussed above. For the calculation, we used the parameters of the Co/Ni multilayer, Si (Sub.)/Ta 4 /Pt 2 /Co 0.3 /Ni 0.6 / Co 0.3 / MgO 1 /Pt 2/ Ta 4 (the numbers are in nanometers), as reported in our previous study (see also Table I) [20]. Note that there is a threshold $H_x$ at which it starts affecting on $\sigma_{\text{DW,total}}^2$ (and thus $H_n$), as shown in Fig. 1(b). In addition, it is clearly observed that the initial value of $\sigma_{\text{DW,total}}^2$ at $H_x = 0$ mT depends on $H_{\text{DMI}}$. This indicates that $\sigma_{\text{DW,total}}$ decreases because of the DMI and is independent of $H_x$ up to the threshold $H_x$ owing to the compensation of $H_x$ effect between DW1 and DW2. When $H_x$ becomes larger than the threshold, $\sigma_{\text{DW,total}}$ becomes dependent on $H_x$, as described by Eq. (3). The $[\sigma_{\text{DW,total}}(H_x)]^2/[\sigma_{\text{DW,total}}(H_x = 0)]^2$ vs. $H_x$ plots in Fig. 1(c) clearly show that the threshold $H_x$ is a DMI-dependent characteristic; it increases linearly with $H_{\text{DMI}}$. Therefore, according to this extended droplet model, $H_{\text{DMI}}$ can be estimated from a threshold $H_x$ above which $H_n$ varies significantly with $H_x$.

To verify the extended droplet model, we experimentally examined the $H_n$ of the Co/Ni multilayer in terms of $H_x$. The Co/Ni multilayer with a Pt underlayer, as mentioned above, was prepared using the same method as that used in Ref. [20]. In this study, the magnetization switching field ($H_{\text{sw}}$) was measured from the AHE under a tilted external field to obtain the values of $H_n$. For the measurement, a 5 μm×90 μm microstrip was fabricated using conventional UV lithography and Ar ion milling. The values of $H_{sw}$ are determined in terms of the polar angle $\theta$ from the AHE curves when the magnetization switches from up to down and *vice versa*, as displayed in Fig. 2(a) and (b), respectively. The measured $H_{sw}$ vs. $\theta$ [$H_{sw}(\theta)$] is plotted in Fig. 2(c). Note that $H_{sw}(\theta)$ follows $1/\cos\theta$ behavior, which is known as the Kondorsky model [28-30]. Therefore, the switching mechanism of the Co/Ni strip is the DW propagation followed by the reversed domain nucleation required in this study. We obtained $H_x$ and $H_n$ from the measured values of $H_{sw}$ using the following relations: $H_x = H_{sw}(\theta)\cdot\sin\theta$ and $H_n = H_{sw}(\theta)\cdot\cos\theta$, where $\theta$ is the angle between the external field and film normal (z). As predicted by the extended droplet model, the plot of $H_n$ vs. $H_x$ exhibits a threshold [see Fig. 2(d) and its inset]. Here, the values of $H_n$ are normalized by $H_{sw}$ ($H_x = 0$) to eliminate the other parameters in Eq. (1), except $\sigma_{DW,total}$, i.e., $\sigma_{DW,total}$ is the only parameter that depends on $H_x$. $H_{DMI}$ and $D$ are confirmed to be 228±60 mT and 0.45±0.15 mJ/m$^2$, respectively, using the best fitting. These values are almost consistent with the DW velocity measurement results listed in Table I [20]. Therefore, the extended droplet model quantitatively explains the effect of the DMI on magnetic domain nucleation.

To confirm the validity of the extended droplet model for the estimation of the DMI strength, we calculate an energy barrier ($E_B$) as a function of $H_x$ and an out-of-plane field ($H_z$), based on the string method [31]. The energy barrier is calculated by initially setting a transition path between two minima (approximately all up or down spins), which is followed by obtaining discrete images of the path (image number i=0, …, 100). Then, we update the initial path, which is not the minimum energy path (MEP), *via* the damping term of the Landau–Lifshitz–Gilbert equation, $\hat{\mathbf{m}}_i(t+\Delta t) = \hat{\mathbf{m}}_i(t) - \int_{t}^{t+\Delta t} \gamma \mu_0 \alpha \hat{\mathbf{m}} \times (\hat{\mathbf{m}} \times \mathbf{H}) dt$, until it reaches the MEP. Here, $\gamma$ is the gyromagnetic ratio, $\alpha$ (=1, i.e., overdamping) is the damping constant, and $H$ is the effective field including exchange, anisotropy, magnetostatic, DMI, and external fields. For the simulation, the following parameters are selected from those listed in Table 1: an exchange stiffness constant of $A$ = 6.4 pJ/m, a perpendicular anisotropy energy density of $K_u$ = 1.55 × 10$^6$ J/m$^3$), and a saturation magnetization, $M_s$, of 837 kA/m. We varied the DMI constant, $D$, from 0.3–0.75 mJ/m$^2$. We perform the reparametrization step for the simulation through linear interpolation at every 100 iteration steps, to keep the images equidistant in the phase space [32].

The string model finds the MEP and determines the energy saddle point between two energy minima corresponding to the initial and final states [Fig. 3(a)]. Then, the energy barrier, $E_B$, is given by the difference between the energies at the minimum point (the initial state) and the saddle point. Figure 3(b) shows the contour diagram of $E_B$ as a function of $H_x$ and $H_z/H_{sw}$ ($H_x = 0$) for various DMI constants. Here, we normalize the vertical axis with $H_{sw}$ when $H_x = 0$, assuming the same value of $E_B$ as that in the experiment.

Regardless of the DMI constant, we find that the energy barrier varies significantly across a clear boundary. As $H_x$ increases across the boundary, the magnetic configuration of the droplet at the energy saddle point changes considerably. On the left side of the boundary, the droplet is of the Néel type [Fig. 3(c)], corresponding to "L" in Fig. 3(b), whereas on the right side of the boundary, all in-plane components of magnetizations are aligned along the $x$ direction [Fig. 3(d)], corresponding to "R" in Fig. 3(b). This indicates that on the right side of the boundary, $H_{DMI}$ is completely compensated in the $x$ direction. Therefore, $H_{DMI}$ can be approximately estimated from the threshold $H_x$, which is indicated by a black arrow in each panel in Fig. 3(b). As the DMI constant increases, the threshold $H_x$ shifts to the right, which is consistent with the extended droplet model that includes the DMI. The simulation results indicate that a larger $D$ induces a larger threshold $H_x$, which is also indicated by the extended droplet model. By comparing the numerical and experimental results, we estimate a DMI value of $D \cong 0.45$ mJ/m$^2$, which justifies the extended droplet model.

In conclusion, we developed an extended droplet model considering the energy difference between two DWs with opposite magnetization directions in a droplet under an in-plane field. The model describes that the $\sigma_{\text{DW,total}}$ of the droplet is constant under $H_x < H_{\text{DMI}}$, while $\sigma_{\text{DW,total}}$ becomes dependent on $H_x$ when $H_x > H_{\text{DMI}}$, resulting in the threshold point. The experimental results clearly demonstrate the threshold $H_n$. The value of $D$ obtained from the $H_n$ vs. $H_x$ curve of the asymmetric Co/Ni microstrip is consistent with the values obtained from DW measurement and the full numerical string model. The string model simulation demonstrates that the compensation of $H_{\text{DMI}}$ generated at a counter DW in the droplet results in the threshold $H_n$. The results of this study provide a simple method to describe the DMI effect on the magnetization reversal of a perpendicularly magnetized droplet and a simple electrical measurement method to quantitatively determine the DMI energy density, which is a key factor for next generation spin-orbitronic devices.

Reference

1. A. N. Bogdanov and U. K. Rößler, Chiral symmetry breaking in magnetic thin films and multilayers. *Phys. Rev. Lett.* **87**, 037203 (2001).

2. N. Romming, C. Hanneken, M. Menzel, J. E. Bickel, B. Wolter, K. von Bergmann, A. Kubetzka and R. Wiesendanger, Writing and deleting single magnetic skyrmions. *Science* **341**, 636 (2013).

3. M. Bode, M. Heide, K. von Bergmann, P. Ferriani, S. Heinze, G. Bihlmayer, A. Kubetzka, O. Pietzsch, S. Blügel and R. Wiesendanger, Chiral magnetic order at surfaces driven by inversion asymmetry. *Nature* **447**, 190 (2007).

4. A. Fert, V. Cros and J. Sampaio, Skyrmions on the track. *Nat. Nanotech.* **8**, 152 (2013).

5. A. Hoffmann and S. D. Bader, Opportunities at the Frontiers of Spintronics. *Physical Review Applied* **4**, 047001 (2015).

6. K. –W. Moon, D. –H. Kim, S. –C. Yoo, S. –Ge. Je, B. S. Chun, W. Kim, B. –C. Min, C. Hwang and S. –B. Choe, Magnetic bubblecade memory based on chiral domain walls. *Scientific reports* **5**, 9166 (2015).

7. A. Soumyanarayanan, N. Reyren, A. Fert, and C. Panagopoulos, Emergent phenomena induced by spin-orbit coupling at surfaces and intefaces. *Nature* **539**, 509 (2016).

8. I. E. Dzyaloshinskii, Thermodynamic theory of weak ferromagnetism in antiferromagnetic substances. *Sov. Phys. JETP* **5**, 1259 (1957).

9. T. Moriy, Anisotropic superexchange interaction and weak ferromagnetism. *Phys. Rev*. **120**, 91–98 (1960).

10. A. Fert, Magnetic and transport properties of metallic multilayers. *Mater. Sci. Forum* **59**, 439 (1990).

11. C. Moreau-Luchaire, C. Moutafis, N. Reyren, J. Sampaio, C. A. F. Vaz, N. Van Horne, K. ouzehouane, K. Garcia, C. Deranlot, P. Warnicke, P. Wohlhüter, J.-M. George, M. Weigand,J. Raabe, V. Cros and A. Fert, Additive interfacial chiral interaction in multilayers for stabilization of small individual skyrmions at room temperature. *Nature nanotechnology* **11**, 444 (2016).

12. Olivier Boulle, Jan Vogel, Hongxin Yang, Stefania Pizzini, Dayane de Souza Chaves, Andrea ocatelli, Tevfik Onur Menteş, Alessandro Sala, Liliana D. Buda-Prejbeanu, Olivier Klein,Mohamed Belmeguenai, Yves Roussigné, Andrey Stashkevich, Salim Mourad Chérif, Lucia Aballe, Michael Foerster, Mairbek Chshiev, Stéphane Auffret, Ioan Mihai Miron & Gilles Gaudin, Room-temperature chiral magnetic skyrmions in ultrathin magnetic nanostructures. *Nat. Nanotech.* **11**, 449 (2016).

13. K. Litzius, I. Lemesh, B. Krüger, P. Bassirian, L. Caretta, K. Richter, F. Büttner, Koji Sato, Oleg A. Tretiakov, Johannes Förster, Robert M. Reeve, MarkusWeigand, Iuliia Bykova, Hermann Stoll, Gisela Schütz, G. S. D. Beach and Mathias Kläui, Skyrmion Hall effect revealed by direct time-resolved X-ray microscopy. *Nat. Phys.* (2016): DOI: 10.1038/NPHYS4000.

14. S. Woo, K. Litzius, B. Krüger, M.-Y. Im, L. Caretta, K. Richter, M. Mann, A. Krone, R. M. Reeve, M. Weigand, P. Agrawal, I. Lemesh, M. –A. Mawass, P. Fischer, M. Kläui and G. S. D. Beach, Observation of room temperature magnetic skyrmions and their current-driven dynamics in ultrathin Co films. *Nat. Mater.* **15**, 501-506 (2016).

15. S. Emori, U. Bauer, S. –M. Ahn, E. Martinez and G. S. D. Beach, Current-driven dynamics of chiral ferromagnetic domain walls. *Nat. mater.* **12**, 611 (2013).

16. K. –S. Ryu, L. Thomas, S. –H. Yang and S. Parkin, Chiral spin torque at magnetic domain walls. *Nat. nanotech.* **8**, 527 (2013).

17. K. Ueda, K.-J. Kim, T. Taniguchi, T. Tono, T. Moriyama, and T. Ono, In-plane field-driven crossover in the spin-torque mechanism acting on magnetic domain walls in Co/Ni. *Physical Review B* **91**, 060405(R) (2015).

18. J. Torrejon, J. Kim, J. Sinha, S. Mitani, M. Hayashi, M. Yamanouchi and H. Ohno, Interface control of the magnetic chirality in CoFeB/MgO heterostructures with heavy-metal underlayers. *Nat. commun.* **5**, 4655 (2014).

19. S. -H. Yang, K. –S. Ryu and S. Parkin, Domain-wall velocities of up to 750 m s$^{-1}$ driven by exchange-coupling torque in synthetic antiferromagnets. *Nat. Nanothech.* **10**, 221 (2016).

20. Y. Yoshimura, K.-J. Kim, T. Taniguchi, T. Tono, K. Ueda, R. Hiramatsu, T. Moriyama, K. Yamada, Y. Nakatani, and T. Ono, Soliton-like magnetic domain wall motion induced by the interfacial Dzyaloshinskii–Moriya interaction. *Nat. Phys.* **12**, 157 (2016).

21. R. Tomasello, M. Carpentieri, and G. Finocchio, Influence of the Dzyaloshinskii-Moriya interaction on the spin-torque diode effect *J. Appl. Phys.* **115**, 17C730 (2014).

22. N. Perez, E. Martinez, L. Torres, S.-H. Woo, S. Emori, and G. S. D. Beach, Chiral magnetization textures stabilized by the Dzyaloshinskii-Moriya interaction during spin-orbit torque switching *Appl. Phys. Lett.* **104**, 092403 (2014).

23. P.-H. Jang, K. Song, S.-J. Lee, S.-W. Lee, and K.-J. Lee, *Appl. Phys. Lett*. **107**, 202401 (2015).

24. J. Sampaio, A. V. Khvalkovskiy, M. Kuteifan, M. Cubukcu, D. Apalkov, V. Lomakin, V. Cros, and N. Reyren, *Appl. Phys. Lett*. **108**, 112403 (2016).

25. S. Pizzini, J. Vogel, S. Rohart, L. D. Buda-Prejbeanu, E. Jué, O. Boulle, I. M. Miron, C. K. Safeer, S. Auffret, G. Gaudin, and A. Thiaville, Chirality-Induced Asymmetric Magnetic Nucleation in Pt/Co/AlO$_x$ Ultrathin Microstructures. *Phys. Rev. Lett.* **113**, 047203 (2014).

26. J. Moritz, B. Dieny, J. P. Nozières, Y. Pennec, J. Camarero, and S. Pizzini, *Phys. Rev. B* **71**, 100402(R) (2005).

27. S. –G. Je, D. –H. Kim, S. –C. Yoo, B. –C. Min, K. -J. Lee, and S.-B. Choe, Asymmetric magnetic domain-wall motion by the Dzyaloshinskii-Moriya interaction. *Phys Rev B* **88**, 214401 (2013).

28. E. Kondorsky, A mechanism of magnetic hysteresis in heterogeneous alloys. *J. Phys.* (Moscow) **2**, 161 (1940).

29. F. Schumacher, On the modification of the Kondorsky function. *J. Appl. Phys.* **70**, 3184 (1991).

30. K. R. Coffey, T. Thomson, and J. –U. Thiele Angular dependence of the switching field of thin-film longitudinal and perpendicular magnetic recording media. *J. Appl. Phys.* **92**, 4553 (2002).

31. E. Weinan, W. Ren, and E. Vanden-Eijnden, Simplified and improved string method for computing the minimum energy paths in barrier-crossing events. *J. Chem. Phys.* **126**, 164103 (2007).

32. Gabriel D. Chaves-O'Flynn, Daniel Bedau, Eric Vanden-Eijnden, Andrew D. Kent and Daniel L. Stein. Stability of 2π Domain Walls in Ferromagnetic Nanorings. *IEEE Trans. Magn.* **46**, 2272 (2010).

Acknowledgement

This work was supported by JSPS KAKENHI Grant Numbers 15H05702, 26870300, 26870304, 26103002, Collaborative Research Program of the Institute for Chemical Research, Kyoto University, and R & D Project for ICT Key Technology of MEXT from the Japan Society for the Promotion of Science (JSPS). D.K. was supported from Overseas Researcher under Postdoctoral Fellowship of Japan Society for the Promotion of Science (Grant Number P16314). KJL acknowledges the National Research Foundation of Korea (NRF) grant funded by the Korea government (MSIP) (2015M3D1A1070465).

Table I. Physical parameters of the Co/Ni microstrip obtained in this study and ref. [20].

| Method | $\mu_0 M_S$ [A/m] | $\Delta$ (nm) | $\sigma_0$ (mJ/m$^2$) | $K_D$ (10$^4$ J/m$^3$) | $H_{DMI}$ (mT) | $D$ (mJ/m$^2$) |
|---|---|---|---|---|---|---|
| Nucleation (this study) | 0.837 × 10$^6$ | 2.4±0.2 | 10.7±1.0 | 3.6±1.6 | 228±60 | 0.45±0.15 |
| DW velocity (Ref. [20]) | | 3.4 | 11.9 | 2.1 | 213±55 | 0.60±0.15 |

Figures

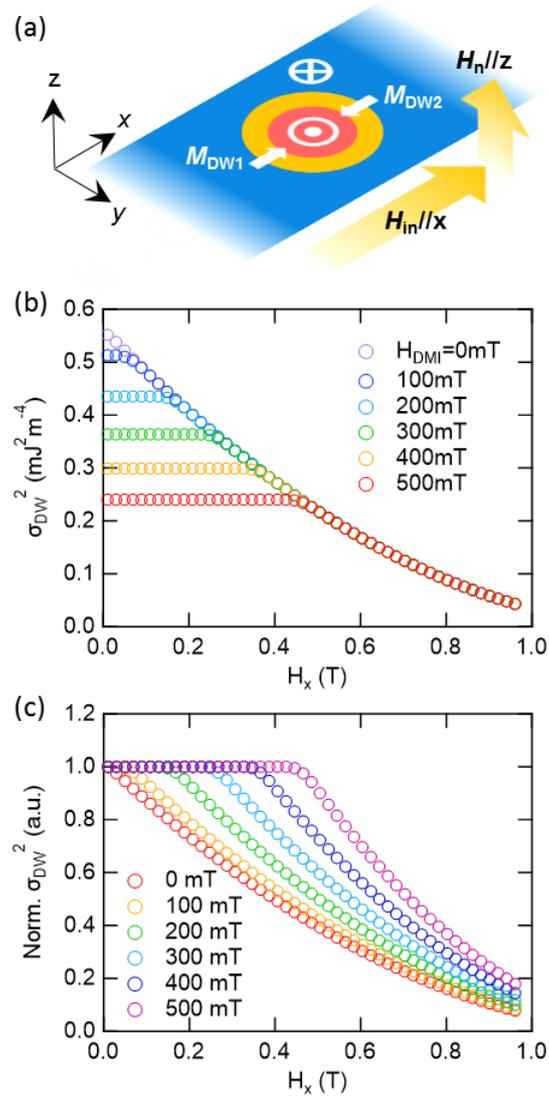

Figure 1. (a) Schematic image of the magnetic droplet (red part in the strip) in the perpendicularly magnetized system (blue part) with opposite magnetization directions. Yellow border indicates the DW. The magnetization of the DW is illustrated with white arrows. The in-plain ($H_x$) and out-of-plane nucleation fields ($H_n$) are illustrated using yellow arrows. (b) $\sigma_{DW,total}^2$ vs. $H_x$ and (c) normalized $\sigma_{DW,total}^2$ vs. $H_x$ curves in terms of selected $H_{DMI}$ values. The method of normalization is explained in the text.

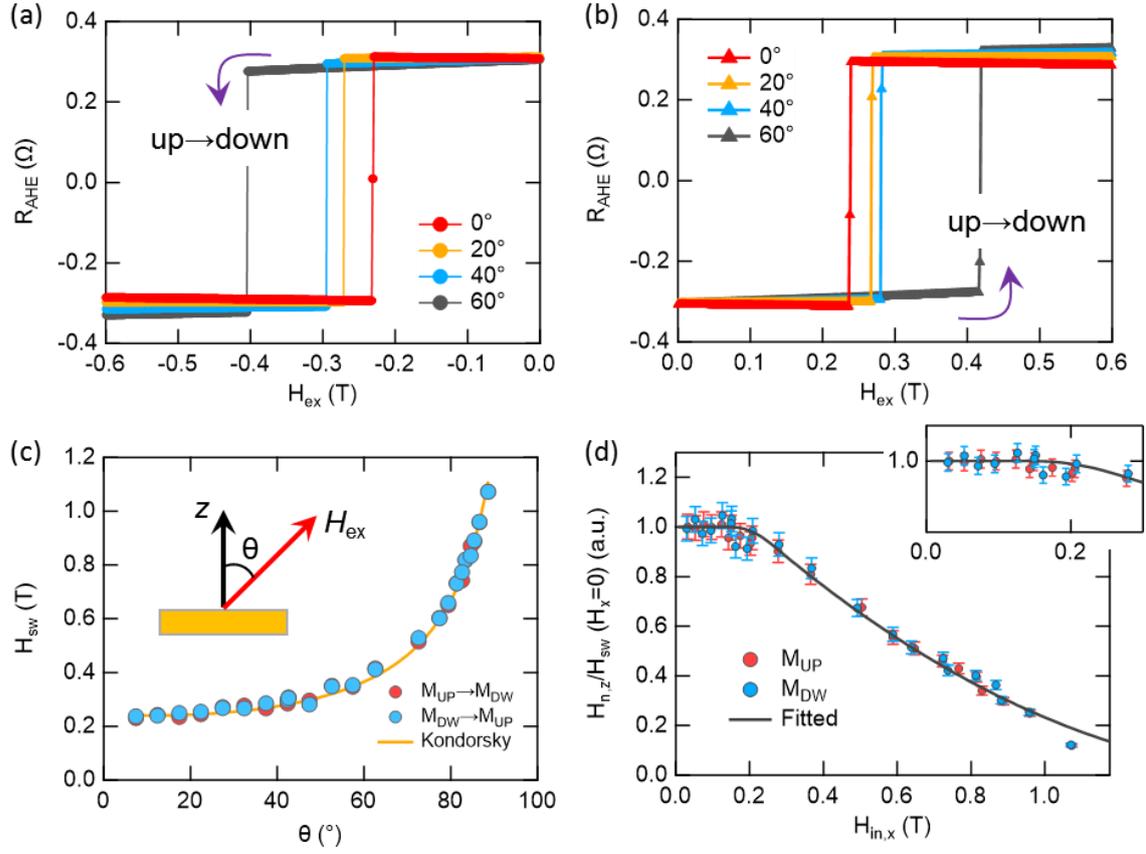

Figure 2. AHE curves when magnetization switches (a) from up to down, and (b) from down to up as a function of the magnetic field angle (θ) with respect to the film normal [see the inset of (c)]. (c) $H_{sw}$ plots in terms of θ. The inset illustrates the field angle and film normal. The yellow line shows the $H_{sw}(θ) = H_{sw}(0)/\cos θ$ curve, which is known as the Kondorsky model. (d) $H_n/H_c$ ($H_x = 0$) plots in terms of $H_x$ with up and down magnetized states. The black solid lines show the fitting result obtained using the extended droplet model. The inset shows the zoomed-in image of the curve in a field range 0–300 mT.

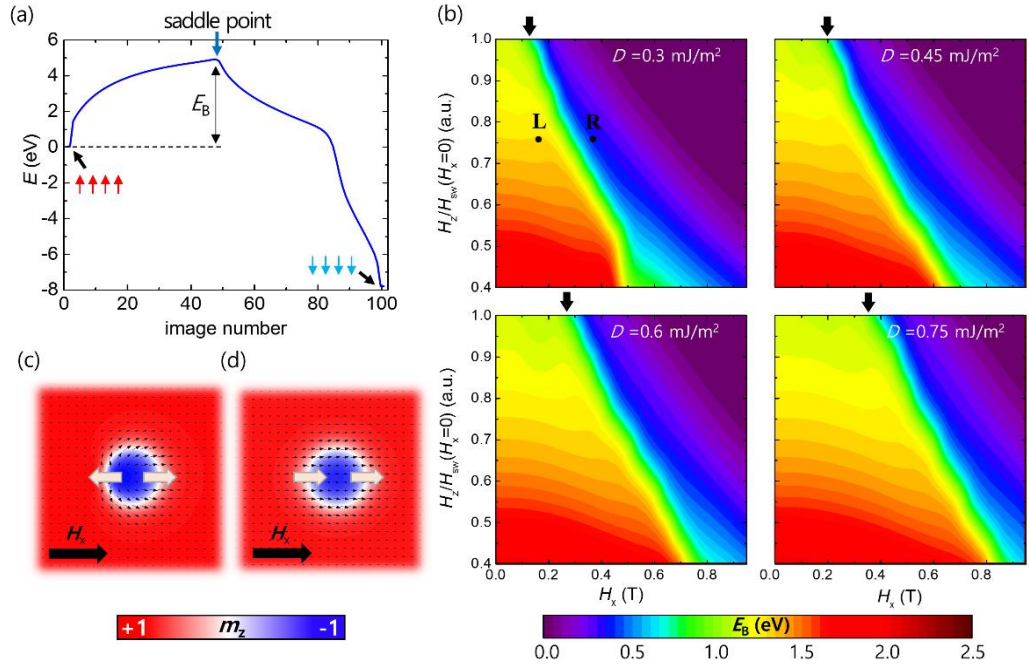

Figure 3. (a) Energy barrier of droplet. (b) Contour diagram of $E_B$ as a function of $H_x$ and $H_z$ for the following values of $D$: 0.3, 0.45, 0.6, 0.75 mJ/m$^2$. Black arrows in each panel correspond to the threshold $H_x$. Magnetization configuration at a saddle point for (c) $|H_x| < |H_{DMI}|$, corresponding to "L" in (b) and (d) $|H_x| > |H_{DMI}|$, corresponding to "R" in (b). The color code corresponds to the out-of-plane component of magnetization and the arrows correspond to the in-plane component of magnetization.